\documentclass[acmsmall]{acmart}
\usepackage{enumerate}
\usepackage{multirow}
\usepackage[normalem]{ulem}
\useunder{\uline}{\ul}{}
\usepackage{makecell}
\usepackage{url} 
\usepackage{hyperref}
\usepackage[hyphenbreaks]{breakurl}

\AtBeginDocument{%
  }

\setcopyright{rightsretained}
\acmJournal{PACMHCI}
\acmYear{2023} \acmVolume{7} \acmNumber{CSCW2} 
\acmArticle{301} \acmMonth{10} \acmPrice{}
\acmDOI{10.1145/3610092}

\acmPrice{15.00}
\acmISBN{978-1-4503-XXXX-X/18/06}

\received{January 2023}
\received[revised]{April 2023}
\received[accepted]{July 2023}

\newcommand{\add}[1]{\textcolor{red}{#1}} 
\newcommand{\del}[1]{\add{\st{#1}}}

\renewcommand{\add}[1]{#1}
\renewcommand{\del}[1]{}

\makeatletter
\gdef\@copyrightpermission{
  \begin{minipage}{0.2\columnwidth}
   \href{https://creativecommons.org/licenses/by/4.0/}{\includegraphics[width=0.90\textwidth]{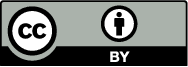}}
  \end{minipage}\hfill
  \begin{minipage}{0.8\columnwidth}
   \href{https://creativecommons.org/licenses/by/4.0/}{This work is licensed under a Creative Commons Attribution International 4.0 License.}
  \end{minipage}
  \vspace{5pt}
}
\makeatother

\begin{document}
\title[Moderation Strategies in OSS Projects and the Role of Bots]{\textit{``Nip it in the Bud"}: Moderation Strategies in Open Source Software Projects and the Role of Bots}

\author{Jane Hsieh}
\email{jhsieh2@cs.cmu.edu}
\affiliation{%
  \institution{Carnegie Mellon University}
  \country{USA}
}

\author{Joselyn Kim}
\email{joselynk@andrew.cmu.edu}
\affiliation{%
  \institution{Carnegie Mellon University}
  \country{USA}
}

\author{Laura Dabbish}
\email{dabbish@cs.cmu.edu}
\affiliation{%
  \institution{Carnegie Mellon University}
  \country{USA}
}

\author{Haiyi Zhu}
\email{haiyiz@cs.cmu.edu}
\affiliation{%
  \institution{Carnegie Mellon University}
  \country{USA}
}
\renewcommand{\shortauthors}{Jane Hsieh et al.}

\begin{abstract}
Much of our modern digital infrastructure relies critically upon open sourced software. The communities responsible for building this cyberinfrastructure require maintenance and moderation, which is often supported by volunteer efforts. Moderation, as a non-technical form of labor, is a necessary but often overlooked task that maintainers undertake to sustain the community around an OSS project. This study examines the various structures and norms that support community moderation, describes the strategies moderators use to mitigate conflicts, and assesses how bots can play a role in assisting these processes. We interviewed 14 practitioners to uncover existing moderation practices and ways that automation can provide assistance. Our main contributions include a characterization of moderated content in OSS projects, moderation techniques, as well as perceptions of and recommendations for improving the automation of moderation tasks. We hope that these findings will inform the implementation of more effective moderation practices in open source communities.
\end{abstract}



\maketitle

\section{Introduction}
Online social coding platforms such as GitHub facilitate the production of open source software (OSS), which modern digital infrastructure relies heavily upon. However, excess volumes of issues and requests filed by users can overload volunteer project maintainers \cite{Raman2020-rd}. Aggravating the situation, open source developers can become toxic and hostile in the course of technical or ideological disagreements \cite{Cohen2021-na}. Incivility suppresses productivity, creativity and quality in the workplace \cite{Porath2013-mg}, and for semi-professional software production platforms like GitHub, such misbehaviors have caused growing concerns over the mental well-being of contributors and maintainers \cite{Mancini_undated-dg}. 

Moderation, as a non-technical form of labor, is a necessary but often overlooked \add{and understudied} task that maintainers undertake to sustain the community around an OSS project. 
\del{Previous work in open source has not considered in depth}
\add{To date, it is not well understood} 
how maintainers grapple with toxic or undesirable behavior on their projects, particularly at scale. 
Research has described the different types of conversations around code contributions \cite{Tsay2014-nr} and categorized toxic content as insults, trolling, as well as displays of arrogance and entitlement \cite{Cohen2021-na, Miller_undated-vr}. At the same time we know that responding to issues and pull requests is an important part of maintenance work in open source \cite{Eghbal2020-lp,Geiger2021-ay}. As Geiger points out, maintainers must delicately navigate instances where there is a mismatch between the work required to merge a contribution and non-maintainers' desires to integrate a certain piece of functionality \cite{Geiger2021-ay}. We also know that responses and interactions around public conversations on OSS projects in GitHub are an important signal to potential contributors and users of a project, underscoring the importance of dealing with toxicity \cite{Qiu2019-xt,Dabbish2012-yi}.

A growing body of research in CSCW examines how users moderate their own content in online communities and increasingly leverage automation to more efficiently control bad behavior \cite{Seering2019-hm,Lampe2004-pg,Wang2014-ue}. These studies describe the challenges of moderation in different platforms (e.g. \cite{Jhaver2018-ln,Jiang2019-tr}), explores novel moderation techniques and tools (e.g. \cite{Chandrasekharan2019-aw,Chandrasekharan2022-xi}) and examine the effectiveness of different moderation behaviors and strategies (e.g. \cite{Seering2019-hm,Seering2017-kj}). For example, Jhaver et al. find that moderation transparency matters - offering removal explanations on Reddit reduces the likelihood of future post removals \cite{Jhaver2019-jm}. In \cite{Lampe2004-pg}, Lampe and Resnick observe that timeliness trades off with accuracy in distributed moderation systems like Slashdot. And while automated moderation systems scale well in removing obviously undesirable content (e.g. spam and malware links), Chancellor et al. note how they can magnify errors \cite{Chancellor2017-ib}, making human decisions more preferable for nuanced \cite{Chancellor2018-nz,Jhaver2019-zl} and high stakes contexts \cite{Gray2020-es}. 

But the online social media communities and platforms studied in much of the moderation research vary in three important ways from open source software development. First, social media forums and text-based discussion groups are typically informal public spaces where people gather to share compelling and interesting information, converse, as well as build communities \cite{oldenburg1999great}, whereas open source communities aim to collaboratively produce software, which can entail complex organizational structures and highly technical discussions tied with code artifacts and software that is utilized for professional purposes \cite{bird2008latent}. 
Secondly, each individual contributor's activities on GitHub have implications for employment prospects and reputation, both within OSS and the professional community more broadly \cite{resume}. Unlike their peers on discussion groups like Reddit, where participants are pseudonymous or anonymous, \footnote{choices that may lead to disinhibition online \cite{Suler2004-se,Lapidot-Lefler2012-dw,Yun2020-km}} a large portion of GitHub users are real name identified, and often their accounts are listed on personal CVs or resumes \cite{silvestri2015linking,dabbish2012leveraging}. 
Finally, the types of inappropriate behaviors and harmful content that present in OSS communities diverge from what is traditionally found in social media. Past work has uncovered that passive-aggressive behaviors such as name-calling and entitlement are more prevalent among conversations between OSS developers \cite{Ferreira2021-id, Miller_undated-vr}, and our findings support these results. 
Thus, the distinctive user goals, behaviors and inappropriate content found in OSS communities might necessitate the adoption of unconventional moderation strategies.

In this study, we qualitatively examine community moderation in open source repositories, that is existing strategies, structures and techniques used for mitigating and preventing inappropriate activity and conversation. Moderation here includes activities to manage behavior in conversations around issues and code contributions as well as the code itself (e.g. the use of potentially offensive variable names). Specifically, we sought to answer the following research questions to investigate moderation in open source communities:

\textbf{Research questions: }
\begin{enumerate}
    \item What does moderation look like in OSS?
    \begin{enumerate} 
        \item Who performs moderation actions in projects and in what capacity
        \item What strategies do moderators use to respond to, diffuse and prevent conflicts?
\end{enumerate}
    \item What are the current limitations of automation for moderation and potential future improvements?
\end{enumerate}

In order to address these questions, we conducted interviews with 14 maintainers across 10 projects to identify how moderation actions are performed on projects of different scales as well as attitudes towards algorithmic support for toxicity moderation and prevention. We find that 1.) moderation in open source is conducted by different roles depending on the size and structure of projects and 2.) moderators leverage several strategies to mitigate and prevent emergent conflicts and 3.) that future efforts will need to address concerns around customizability and detection accuracy before deploying automation tools to help offload the labor of moderation. By documenting the structures and forms of labor performed around moderation within open source projects, we hope to enlighten future practitioners on the available strategies for moderating digitally-mediated software development contexts. By characterizing the potentials and limitations of automation tools for moderation, we support practitioners in understanding and anticipating the challenges and impacts of adopting such automation. We also encourage tool designers and developers to build on these findings, so that future tools for moderation can provide improved and wider services to open source community members.

\section{Background}
Software development is a product-oriented and collaborative endeavor, making open source development environments semi-formal working spaces that expect professional conduct from their participants. During the development process, project collaborators may encounter a myriad of technical and interpersonal conflicts that impede their work. In the following we present our study platform, notable types of open source conflicts that prior work and our participants have reported, as well as relevant prior work on moderation and automation for toxicity detection in various online communities.

\subsection{Study Context: GitHub}
We focused our data collection on open source software projects hosted on the GitHub platform. GitHub facilitates collaboration and communication among developers, users and owners of software projects \cite{Dabbish2012-yi,Li2021-rx}, and is arguably the most popular hosting site for projects. As of June 2022, GitHub reports having over 83 million developers \cite{users} and more than 200 million repositories (including at least 28 million public repositories).

Projects on GitHub are organized into code \textbf{repositories} (or repos for short) which can be owned by a personal account (usually the creator or another maintainer) or by an organization, which comprises multiple users. Collaborators to the repository have direct write access to make commits, and they work together with the owner to maintain the project. Users primarily consist of software consumers, and they can star repos to express interest in a project or to save it for later reference. Within each repository, contributors and users can plan work, track bugs, request new features or express maintenance concerns by creating \textbf{issues} \cite{Li2021-rx}. When an external (non-collaborator) developer has changes to propose, they can submit a \textbf{pull request} -- a special issue for posting code contributions so that others can review and integrate them into the existing codebase. However, a pull request requires approval by one or more authorized collaborators before it can be merged. To communicate about developments, collaborators can comment under issues, pull requests as well as lines of code.

\subsection{Conflicts and Incivility in Open Source} \label{conflicts}
Open source project maintainers are responsible for tremendous amounts of unseen civic labor that underlies our digital infrastructure, and many have documented how overwhelming volumes of such invisible labor engenders harm to the mental well-being \cite{noauthor_2017-bz,Eghbal2016-if}. Maintainers are seldom recognized sufficiently for their stewardship, causing individual stress and burnout \cite{Raman2020-rd}, imperiling projects for undermaintenance, and threatening the overall sustainability of the open source ecosystem \cite{Cohen2021-na, Raman2020-rd}. 
Due to factors like a lack of corporate management structure and geographic dispersion, open source maintainers are required to undertake a plethora complex interpersonal and organizational work \cite{Geiger2021-ay}. Community maintenance tasks include providing support to internal contributors as well as technical assistance to external users so they can make use of the product. Previous investigations have found such organizational and interpersonal labor to play a critical role in traditional software engineering contexts \cite{nagappan2008influence, matturro2013soft, tamburri2013organizational}. Due to the fully public and largely voluntary nature of discussions and actions in open source development, moderation is one necessary task that maintainers must undertake to avoid an overwhelming amount of negative content and harmful interactions.  

Prior work extensively documented the presence of incivility, conflict and in general negative emotions across multiple actions of open source development, including code reviews \cite{Ferreira2021-id,Egelman2020-xs,Bosu2013-wi,Ranzhin_undated-cd,Dietrich_undated-zz,Arneja2015-lx,Ahmed2017-wt}, issue discussions \cite{Ferreira2022-cc,Miller_undated-vr} as well as in comments to these actions \cite{Gachechiladze2017-op,Guzman2014-po}. Negative interactions occur among different members of the community (e.g. core collaborators, external contributors as well as maintainers across different projects), and stem from multiple grounds, ranging from language and cultural differences to political disagreements to personal feuds to software dependencies to mismatches in expectations \cite{Li2021-rx,Filippova2015-yi,Geiger2021-ay}. Conflicts among internal contributors can be difficult to moderate - since organization members cannot ban each other and interventions between familiar and respected contributors can get tricky - but politically charged misconduct from external and banned members can be harmful as well. Incivil behaviors present in the semi-professional volunteer software development environment endanger the sustainability of open source by decreasing the intrinsic motivation of contributors, reducing their productivity, and heightening dropout rates of newcomers \cite{Kaur2022-hx,Miller2019-ve,Porath2013-mg}. Rather than categorizing the types of conflict in open source \add{(the focus of \cite{Filippova2015-yi,Miller_undated-vr,Cohen2021-na})}, one aim of our study (via RQ1) is to characterize strategies and structures that maintainers use to moderate such incivil situations.

While incivility originating from internal contributors of a project has been well-studied \cite{Ferreira2021-id,Egelman2020-xs,Bosu2013-wi,Arneja2015-lx, Ahmed2017-wt,Tan2019-kw, Ranzhin_undated-cd,Dietrich_undated-zz}, frustration that follows from unrealistic expectations of user support can cause toxic insults \cite{Miller_undated-vr} and entitlement directed at maintainers, demanding their time and attention \cite{Geiger2021-ay}. User support involves providing assistance to consumers of the software who have difficulties making use of it either because of existing defects or the consumer's misunderstanding of some aspect of the software \cite{Hippel_undated-uq}. Swarts identified usability and transparency issues as causes of user needs in open source \cite{Swarts2019-fn}. As projects scale, user support becomes a tedious task, overwhelming maintainers with issues and requests, demanding their time and emotional labor \cite{Geiger2021-ay}. Unlike commercial vendors that generally rely on institutional infrastructures such as paid and dedicated IT or tech support teams, open source software provides informational user support free of charge, via a small group of volunteer users and maintainers \cite{Hippel_undated-uq}.

\subsection{Governance and Moderation}
The non-technical labor of moderation is often overlooked but essential for understanding the infrastructure of open source \cite{Lee2006-ln,Ribes2013-kw,Eghbal2016-if}. According to Grimmelmann, moderation consisted of \textit{``governance mechanisms that structure participation in a community to facilitate cooperation and prevent abuse''}  \cite{Grimmelmann_undated-vj}. For the context of open source, we define community moderation to be the set of activities that maintainers and designated moderators leverage to manage behavior in conversations around issues, code contributions, and code itself in an effort to minimize harmful and abusive activities and to foster a collaborative and welcoming environment for contributors.

Much like their social media (e.g. Discord \cite{Jiang2019-tr}, Reddit \cite{Juneja2020-kd,Chandrasekharan2017-qu}, Twitter\cite{jhaver2018online}) and peer production contemporaries (e.g. Wikipedia \cite{Geiger2010-zh,Forte2009-ze,Billings2010-cy}), GitHub communities engage in volunteer-based community moderation, as opposed to platform-wide commercial moderation. \footnote{though GitHub did develop a set of platform-wide Acceptable Use Policies \cite{acceptable}}
Such voluntary nature of moderation and maintenance in open source forces members of the community (e.g., maintainers or volunteer contributors) to bear the responsibility of providing support and assistance to users. But unlike support providers of commercial software products, the services of volunteer contributors are uncompensated \cite{Hippel_undated-uq}. 
To exacerbate their workload, prior studies reported that maintainers found user support to be an \textit{``overwhelming and never-ending chore, particularly for projects that use GitHub-style collaboration platforms''} \cite{Geiger2021-ay}. The staggering volume of demands for user support and feature requests on GitHub's issue-posting mechanisms demonstrates an instance of \textit{overuse} -- a form of deviant behavior among Grimmelmann's categorization of abuses that leads to congestion and cacophony, making it harder for information to get through and thereby hindering users' information search and retrieval processes \cite{Grimmelmann_undated-vj}.



Existing systems of platformic content moderation have been found to vary in terms of actions, styles, philosophies and values. In a systematic review that engaged 86 papers related papers, Jiang et al. described such tradeoffs and compared the various moderation techniques with Grimmelann's four broad categories \cite{Jiang2022-lg}. These included \textit{exclusion} -- the act of depriving people access to the online community, often through bans or timeouts, \textit{organizing} -- consisting of measures like removing and annotating content, \textit{norm-setting} -- a practice of issuing warnings or \textit{``indirect policing''} to denounce bad behavior as well as \textit{monetary pricing} -- a way of using market forces to raise the prices of participation on users -- though social media users were not found to engage with this last category \cite{Grimmelmann_undated-vj, Jiang2022-lg}. In a study of volunteer moderators on Reddit, Facebook and Twitch, Seering et. al. showed how moderators used excluding and norm-setting actions (e.g., bans and warnings) at increasingly restrictive rates and relied heavily on general community members to report and flag misbehaviors \cite{Seering_Wang_Yoon_Kaufman_2019}. While the actions of excluding, organizing and norm-setting may be transferable to open source moderation, we expect that the distinct forms of inappropriate content might motivate the adoption of other unique strategies and governance structures. We sought to characterize the moderation structures, norms and roles involved in open source via RQ2.

While some past work examined conflict management strategies for peer review \cite{Huang2016-mz} and the emergence of early governance structures on GitHub \cite{OMahony2007-zi}, we lack knowledge around the specific strategies that maintainers use to moderate inappropriate and problematic behaviors in open source. Among the many forms of intervention techniques available for such purposes, Renee et al. investigated how the code of conduct, a document that \textit{``defines standards for how to engage in a community\dots signals an inclusive environment that respects all contributions \dots  [and] outlines procedures for addressing problems between members''} \cite{CoC} is used for moderation. Other moderation tools include documents such as contributing guidelines (\textit{``which provides potential project contributors with a short guide to how they can help with your project''}  \cite{Contributing}), moderation policies, or in-house features such as bans, the locking of conversations \cite{locking}. However, Geiger et al. uncovered that contributors are not as intrinsically motivated to engage in non-technical maintenance work (e.g. community support and documentation) as they are to complete more technical tasks (e.g. feature implementation or debugging) \cite{Geiger2018-rx}, indicating a need for more comprehensive and higher-level strategies for conducting moderation for complex situations and interpersonal conflicts. Maintainers can be especially discouraged to perform moderation work since it has been found to cause psychological and emotional distress \cite{Steiger2021-qy}, and automated assistance to moderation can be an appealing solution, with the potential to minimize maintainers' time and labor on tedious tasks and increasing developer productivity \cite{Erlenhov2020-ls,Wessel2020-pm}. However, there exists a gap in our understanding of how OSS moderation is executed in practice, both in terms of strategies as well as the roles and structures that are established to support and facilitate moderation. In this study, we qualitatively investigate such infrastructures and approaches, as well as uncover maintainers' perspectives on how automation can support moderation. 

\subsection{Automated Moderation Bots in Open Source}

Sentiment Bot and Safe Space are examples of tools that leverage existing sentiment analysis models to help maintainers detect and regulate the existence of toxic comments on GitHub. The Sentiment Bot is a GitHub App built with GitHub's Probot framework that \textit{``replies to toxic comments with a maintainer designated reply and a link to the repo's code of conduct''} \cite{sentiment} while Safe Space is a GitHub action that leverages TensorFlow's toxicity classification model to \textit{``detect potential toxic comments added to PRs and issues so authors can have a chance to edit them and keep repos a safe space''} \cite{safespace}. Both of these bots make use of machine learning classifiers to detect toxic content within pull request or issue threads and respond back with a comment that urges the original author to modify or delete their comment whenever problematic content is detected. Underlying such tools are sentiment analysis detectors, and numerous models have emerged in the field of software engineering to improve accuracy and domain specificity of such models. These include classifiers of negative interactions trained on conversations surrounding issues \cite{Raman2020-rd,Sarker2020-ud,Jongeling2017-qe,Gachechiladze2017-op,Murgia2014-fm}, code reviews \cite{Ferreira2022-cc,Bosu2013-wi,Ahmed2017-wt}, commits \cite{Guzman2014-po,Guzman2013-qp}, codes of conduct \cite{Squire2015-ru} as well as data from other contexts such as IT support \cite{Blaz2016-rv} and Stack Overflow \cite{Calefato2018-br}. 

However, bot use in open source contexts has its own associated challenges. Wessel et al. found that bot-generated noise (in the form of verbosity or excessive/undesirable tasks) causes annoyance for contributors, disrupts their workflow, and creates additional labor for maintainers \cite{Wessel2021-pi}. Meanwhile, Huang et al. discovered how contributors react negatively to automated encouragements \cite{Huang2016-mz}. Outside of open source, Jhaver et al. described how subpar removal explanations provided by bots on Reddit brewed community resentment \cite{Jhaver2019-jm}. 
In voice-based communities like Discord, bots faced challenges in identifying rule violations based on nuances such as tone and accent, despite the widespread adoption of bots to automate features \cite{Jiang2019-tr}.
Jiang et. al. highlighted the tradeoff that while automation help communities achieve moderation at massive scales and with faster turnarounds, human involvement is required to understand contextual nuances, provide clear removal explanations, and conduct negotiations around norms that contribute toward community building \cite{Jiang2022-lg}.
Moderators of the three platforms that Seering et. al. studied also expressed the desire to personally deal with harder, more nuanced situations, despite being content to have automated tools deal with the most egregious and unwanted content --- the authors argue that these desiderata are motivated by moderators' inclination to make context-specific judgments and impact community development \cite{Seering_Wang_Yoon_Kaufman_2019}. Smith et al. identified community values related to the design and usage of machine learning-based predictive tools in content moderation on Wikipedia \cite{Smith2020-cl}. 

In open source, maintainers' and moderators' stances toward automation are likely to differ as open source contributors are more habituated to using tooling for increasing productivity and efficiency, whereas the efficiency of moderation has been found to trade off with quality \cite{Jiang2022-lg, Lampe2004-pg}. The second part of RQ2 aims to provide insights on how well current moderation bots support human maintainers in open source contexts and what improvements are needed to reduce friction and concerns in adoption.

\section{Method}
To learn how maintainers and moderators maintain their communities, we interviewed 14 individuals who moderate or maintain projects of varied sizes ranging from 500 to 87,000 stars on their repos and 30 to 4,000 contributors. Before beginning the interview and recruitment process, we obtained institutional IRB approval and debriefed participants on the type of questions to expect prior to starting the interviews to ensure ethicality. 

\subsection{Recruitment}
Participants were recruited through publicly available information on GitHub. The requirement criteria for the participants were that they had to be 1.) at least the age of 18 years old and 2.) either a current or past maintainer or moderator for a collaborative open source project. We started recruiting participants by emailing owners of repos that used moderation bots. But we soon realized that most of these owners had limited moderation experience, since their bot setup resulted from a forked template project. We expanded to recruiting from repos with designated moderation teams or contributing guidelines using search terms such as ``moderating'' or ``moderation team'' on GitHub, and also conducted snowball sampling by asking participants to refer us to other potential interviewees. If the maintainer’s contact information was public, we requested an interview via email. Of the 40 potential participants we emailed, 14 agreed to take the interview, one of whom was female -- the proportion of women involved in this study is on par with the overall representation of women in open source (which is below 5\%) \cite{trinkenreich2022women}. We concluded the recruiting process when the addition of participants stopped generating new emergent themes -- signaling theoretical saturation \cite{creswell2016qualitative}. Table \ref{tab:tab1} displays a summary of participants' projects, their respective roles, as well as descriptive project information.

\subsection{Interview Protocol}
We started the semi-structured interviews by following a protocol of scripted questions, which included questions about negative and positive interactions, detection and moderation strategies, codes of conduct, and bot use. From each category of questions, our main goal was to learn what strategies maintainers used to respond to negative interactions such as violations of codes of conduct and issues with bot usage (after we introduced the Sentiment Bot). Specifically, we inquired about the responsibilities of moderating members, expected norms and behaviors of a community and whenever applicable, their resolution strategies for disruptive behaviors in the past, and how they set precedents for future incidents. Each interview lasted 30-60 minutes and participants were compensated \$15 for their time via PayPal or a donation to a charity or organization of their choice.

\begin{table*}[t]
\resizebox{\textwidth}{!}{
\begin{tabular}{|l|l|l|l|l|l|}
\hline
\textbf{ID} & \textbf{\shortstack{Project \\ Pseudonyms} } & \textbf{Project Area} & \textbf{Role on Project} & \textbf{\shortstack{\#  \\ Contributors}}  & \textbf{Stars} \\ \hline
P1 & Honeysuckle & Visual diagramming platform & Maintainer/Contributor & \textgreater{}20 & \textgreater{}5k \\ \hline
P2 & Receptive & Differential Privacy Library & Maintainer/Contributor & $\sim$50 & \textgreater{}300 \\ \hline
P3 & \multirow{3}{*}{Apex} & \multirow{3}{*}{Runtime Environment} & \multirow{3}{*}{Moderation Team Members} & \multirow{3}{*}{\textgreater{}3k} & \multirow{3}{*}{\textgreater{}85k} \\ \cline{1-1}
P8 &  &  &  &  &  \\ \cline{1-1}
P13 &  &  &  &  &  \\ \hline
P4 & JaguarAPI & \shortstack{Web framework for \\  building APIs}  & Owner/Founder & $\sim$300 & \textgreater{}40k \\ \hline
P5 & Grunge & Programming Language & Designated Moderator & \textgreater{}3.5k & \textgreater{}65k \\ \hline
P6 & Hyundai & Alternative firmware & Designated Moderator & \textgreater{}200 & \textgreater{}17k \\ \hline
P7 & \multirow{2}{*}{Vessel} & \multirow{2}{*}{Container management} & \multirow{2}{*}{Moderation Team Member} & \multirow{2}{*}{\textgreater{}3k} & \multirow{2}{*}{>87k} \\ \cline{1-1}
P9 &  &  &  &  &  \\ \hline
P10 & \multirow{2}{*}{Silverback} & \multirow{2}{*}{Object Storage} & Owner/Founder & \multirow{2}{*}{\textgreater{}200} & \multirow{2}{*}{\textgreater{}500} \\ \cline{1-1} \cline{4-4}
P11 &  &  & Community Manager &  &  \\ \hline
P12 & Kettle & Flash Player Emulator & Owner/Founder & \textgreater{}90 & \textgreater{}9.5k \\ \hline
P14 & Door & Object Storage & Owner/Founder & \textgreater{}80 & \textgreater{}400 \\ \hline
\end{tabular}}
\caption{Participant Summaries. \textit{Project details extracted to preserve anonymity. All references to projects are by pseudonyms}}
\label{tab:tab1}
\end{table*}

\subsection{Analysis}
Using interview recordings and transcripts, a team of two researchers engaged in a bottom-up, thematic analysis of the interviews. The experience of this team in open source contributions ranges from novice to knowledgeable. We adopted a thematic analysis approach to analyze the transcribed video recordings, and followed a shared open coding session to calibrate coding granularity. The first two authors developed the initial lower level codes for each participant’s data and synched weekly to resolve disagreements. After resolving any disagreements amongst the coders, we conducted a bottom-up affinity diagramming process to iteratively refine and group the resulting 375 unique codes into 32 first-level themes, which were then clustered into four main themes that we present below.

\section{Results}
We start by characterizing types of inappropriate behaviors that moderators observed and monitored, separating the common types of rule violations found in other domains from the more implicit forms of conflicts that emerge from the technical development environment of open source. Next, we describe the types of moderation roles and structures that individuals or groups assume or set up in order to more effectively address and govern misconduct. We then discuss the specific strategies that moderators use to react to, address and prevent misbehavior and incivility. Finally, we summarize maintainers' stance around the adoption of tools to automate moderation, highlighting various concerns such as over-censorship, technical incapabilities, as well as limited customizability.

\subsection{Inappropriate behaviors in OSS}\label{goals}
In most well-studied online communities, intolerable behaviors largely comprised of deliberately abusive and disruptive misconduct such as harassment and hate speech \cite{Jiang2019-tr, Jhaver2018-ln, Seering_Wang_Yoon_Kaufman_2019}. However, in the context of open source, explicitly inappropriate behaviors are accompanied by more subtle acts and borderline behaviors such as miscommunication and resistance against new practices. When we inquired about moderation, many maintainers brought up strategies they used to respond to and mediate miscommunication, as well as ways of organizing and curbing the excessive volume of demands. While prior works categorizing toxic behaviors on GitHub have also uncovered less severe misbehaviors such as technical disagreements and arrogance \cite{Miller_undated-vr}, \del{none have made}\add{we make} the distinction between the clearly disruptive content that are detectable by toxicity classifiers and the more covert forms of incivility that require human judgment to identify.
In the following subsection we outline some of the more disruptive acts of misconduct (e.g., hate speech, snarky humor) as well as more subtle forms of misbehaviors that OSS moderators observed and guarded against, and follow with strategies they leveraged to address these in \ref{2} \nameref{2}.

\subsubsection{Explicitly Aggressive and Disruptive Behaviors}
The first class of misbehaviors consisted of explicitly harmful or ill-intended content. We start off by presenting to misconduct that are obvious (e.g., spam) and egregious (e.g. hate speech, harassment) and follow with examples of more concealed (but still harmful) forms of hostility, which include passive aggressiveness and snarky humor. 
\paragraph{Spam, hate speech and harassment}
\begin{itemize}
        \item Much of P8's job as a moderator consisted of  \textit{``moderating spam users''}, which include instances of a \textit{``bot that's leaving nonsensical comments, opening garbage pull requests that are wasting people's time''}. Even in smaller projects such as Hyundai, \textit{``spammers come with things (political) that doesn't have anything to do with Hyundai, they occur twice a year''} (P6).
        
        \item Hate speech like \textit{``someone coming in and saying} `why are you people so stupid', \textit{or worse than that''} can happen but fortunately \textit{``those are very spotty''} (P8). In one case, 
        \del{after someone was banned from the organization, they}\add{a banned member} threatened to \textit{``send collaborators bombs''}
        \del{. The situation is especially alarming because sometime after}
        \add{and afterwards} \textit{``he got arrested, like by the FBI, because he made bombs in his house''} (P8).
        
        \item While commonsense rules like \textit{``no sexual harassment or no discrimination''} seems obvious, P4 pondered how \textit{``in some cases it has to be very explicitly stated, because the people that violate those things are probably the people that wouldn't guess that''}. 
        \item \del{One participant of this study revealed that an offender went so far as to intimidate collaborators of an organization with the threat sending them bombs, which he did actually make in his house, as the FBI later found.}
\end{itemize}

\paragraph{Passive-aggressiveness \& snarky humor }
    \begin{itemize}
        \item Both destructive and contagious \cite{mirela2019qualitative}, \textit{``passive-aggressive comments''} sadly did present in OSS contexts. They include arrogant \textit{``things like`I have been working for 10 years 20 years and I had never seen a solution like what you're proposing' -- something that is not very exclusively saying what you're proposing is dumb, but \dots kind of implicitly saying you're inexperienced \dots in a very, very hidden way''} (P4) or demeaning insults such as\textit{``can’t you ask an intelligent question''} which P4 reports as content that \textit{``we often get within questions threads''}.
        \item In a similar vein, snarky humor is also advised against because \textit{``it's so easy to offend someone with that''} because \textit{``it's really hard to convey what you mean while being snarky on the internet, where nobody can see your face''} (P5).
    \end{itemize}
    
\paragraph{Entitled demands \& heated complaints:} 
\begin{itemize}
    \item Users and contributors who felt entitled to receive responses can take up a significant amount of maintainers' time -- \textit{``the thing that makes most of the time is the questions, issues''} and \textit{``80\% of the time, or like 90\% it will be just like a feature request or a question or [demand for which] like `I'm never really in the user scope}' '' (P4). 
    \item While some requests are easy to address (e.g., simple questions and feature requests), others can get quite heated: ``[One user complained that] \textit{Hyundai was not good because it wasn't working on their device (person didn't read documentation) \dots It started off aggressive, and ended up with the user complaining the documentation wasn't good enough''} (P6). Ironically, \textit{``in many cases [it] is just like errors in the code of the developer (who's asking) and they didn't realize''} (P4). Yet someone must attend to the issues because \textit{``If you ignore people they get more mad \dots and they act out more and more.''} (P11). And the problem with entitled comments isn't the comment itself,\textit{``it's the knock on effects of that comment \dots other people will see that and think it's okay to behave that way \dots [and] feel more entitled because they've seen entitlement be normalized''} (P3).
\end{itemize}

\subsubsection{Misunderstandings, technical disagreements, and resistance against new practices} \label{subtle}
In contrast to explicitly aggressive misbehaviors, our moderator participants also reported monitoring for more subversive disagreements and misunderstandings that arise from the technical and collaborative nature of OSS projects.

\begin{itemize}
        \item Aside from intentional misconduct, \textit{``many times bad behavior is just misunderstandings''} and \textit{``it boils down to like miscommunication and not understanding the issue like people talking past each other and people getting a little bit heated''} (P9). According to P13, miscommunications occurred frequently: \textit{``If you dig into old threads, you see a lot of them are full of miscommunication and people shouting over each other about who should have had dealt with what''}.
        \item Technical disagreements are easy to surface in development environments, because \textit{``sometimes people simply get riled up, they have an idea of what is right or wrong and someone else has a different idea, which in tech can happen''} (P5). In one instance when people did get heated after \textit{``a disagreement with the licensing of the code''} which was \textit{``from another project library''}, some contributors unfortunately \textit{``felt [the need to use] `accusatory' language''}
        \item Technical projects often need to adopt new pipelines and packages to keep up with recent updates and practices, but sometimes new standards are met with \textit{``resistance initially, usually because of large changes such as build pipelines''} (P2). So first they must \textit{``get through the transition period''} (P3), but \textit{``over time there's acceptance''} (P2), \textit{``and the new norms will just be the way it is, and everyone will be horrified that it used to be worse''} (P3).
    \end{itemize}

\subsection{Moderation Roles and Structures } 
While open source communities were once perceived as decentralized and bazaar-like, emergent governance structures form over time \cite{Kilamo2020-en,Bird2011-hr,OMahony2007-zi}. Maintainers in our sample employed a plethora of strategies to overcome interpersonal and technical challenges of social coding. Depending on the size of the project or organization, maintainers varied their governing structure and strategies. Specific moderation actions were performed by members of the community, a moderation team, or maintainers themselves. The most basic form of moderation involved contributors performing self-censorship. After that, volunteer moderators described how they reported potentially harmful content and actions to maintainers or formal moderation teams. In the following sections we describe how participants in our sample described collaboration between different roles and governing powers to conduct moderation together.
\subsubsection{Self-moderation and Volunteer Moderators} \label{self-moderation}
When a particular individual violated the community rule or norm, \textbf{self-moderation\footnote{Acts of self-moderation erases many public records of accidentally posted harmful content, thus we suspect that the practice is prevalent but often undetected. While Apex was the only project reporting self-moderation, it is also one of the most established OSS projects. Hence we expect self-modertating to appear in other projects as well and encourage future work to explore the detection and frequency of self-initiated moderation.} constituted the first line of defense}. 
Unlike the broader term of community self-moderation that Seering proposes \cite{Seering2020-ro}, we consider self-moderation to be the individual self-corrective action of the author to edit and fix their own content, regardless of who first noticed the questionable content. 
In the case of large projects like Apex, maintainers may instate \textit{``an explicit policy to ask organization members to self-moderate}, such rules that \textit{``allow [maintainers] a way to say: `if you just made a mistake, and you apologize and don't do the behavior again, you'll be fine'\dots in a way that displays those norms for the community''} (P3).

\textbf{Member status} affected who received self-moderation requests -- when the original author was \textit{``not a[n internal] collaborator, then the moderation team can just summarily do what we decide is the best''} (P3). However, when internal organization members exhibit problematic behaviors, \textit{``then the first thing [we are required to do] is to always ask them to self-moderate''} (P3). In requesting self-moderation from contributors, maintainers asked for specific actions like editing or deleting the offensive comment, so as to avoid public shaming directed at the author or other escalations. 

Since social coding platforms like GitHub are working environments for producing software, team members are expected to treat each other with civility. So even if  \textit{``You don't have to like each other \dots you have to be professional''} (P13). 
Therefore, when P13 asked contributors who harbor negative feelings toward each other to\textit{``to self-moderate, \dots they did''} and in general \textit{``people \dots are usually regretful that the comment was hurtful \dots [and] will be eager and happy to self-moderate}.''

But in some cases, uncooperative contributors \textbf{refused to conduct self-moderation}, and one cause of this behavior was a difference in cultures. In one case, P13 asked for self-moderation by posting a request along the lines of \textit{``Hey this comment is perceived as \dots problematic, can you please consider self-moderating it''}. If the recipient is from the US, then they would understand that -- \textit{``you're really telling them to do that''}. But \textit{``in Israeli culture, it's perfectly acceptable for them to say `No, I considered it and I think I have a better understanding than you'''.} When contributors refused to cooperate, moderators escalated to more direct measures to intervene, which we cover in \ref{2}.

To delegate some of their responsibilities, maintainers of more popular libraries such as Apex distributed moderation work by \textbf{relying on community members}: \textit{``most of the time somebody reports it \dots they can surface it [by] say[ing] `hey check out this' in a moderation repo that's private to org members''}. While maintainers would prefer to hide contentious content from contributors and users -  \textit{``in an ideal world, we don't require somebody to report it before we fix it''}, some of it inevitably goes undetected in larger projects:\textit{``there's a scalability thing there''}, and community reporting can serve as a \textit{``a useful filter to prevent all of our time being taken up by hunting down problems''} (P3). 

\subsubsection{Formal Moderation Teams} \label{formal}
Larger and more mature projects designated particular volunteer members from the community to form an \textbf{official moderation team} for the organization. The Apex moderation team, for example, consisted of \textit{``8 to 10 people, 5-6 who are regularly active''} (P13). Moderation team members were self-nominated and the role is not even exclusive to contributors- \textit{``any member who is on the project\dots [can] say  `hey I want to be a moderator', and if nobody objects for seven days, they join the team''}. Team members are recertified annually by a Technical Steering Committee (TSC), which guided and advised the organization with higher-level directives. 

Among the ten projects we interviewed, six had designated moderators, all of whom were \textbf{appointed due to existing demand}. For instance, when P12's moderately-sized \textit{``project first started getting popular''}, he had \textit{``no clue how to moderate''}. But growing attention eventually convinced him to assign moderators: \textit{``people were demanding moderators - so very quickly I had to choose a moderator \dots [and these] moderators [would] tell people to calm down and most people are respectful''}.

Maintainers often \textbf{encouraged contributor interactions with moderators} to help offload their maintenance responsibilities. In one case, P5 of the popular programming language Grunge would tell users \textit{``If you have a question about anything that disturbs you or that you may think has disturbed others, contact the moderators''}. P7 of the mature project Vessel also reports how they often redirect users and contributors to \textit{``talk to a moderator on Slack''} whenever the community had questions and doubts around governance actions such as instituted bans, so that moderators can provide them the appropriate explanations. 
Even for more nascent repos such as Silverback, P11 explicitly \textit{``set up community values that proactively explains to people what the community will look like''}, that way \textit{``if someone is blocked, and they don't know why they were blocked, or they think they should be unblocked, they know where to get in touch with us.''} (P8).

Outside of moderating, members may be responsible for \textbf{onboarding} \textbf{tasks} such as taking a training course: \textit{``It was an online course, we went on to Zoom (the whole team) for like a few weeks and we did a training. We should probably do another round [of refreshers] because some people joined} '' (P13). And since the labor of performing moderation actions (e.g. providing explanations) is can be draining \cite{Jiang2022-lg}, it is a moderator's own responsibility to self-assess and take breaks to avoid burnout: \textit{``Moderation is something I do for a while, I stop doing it for a while, I do for a while, I stop doing it for a while, 'cause I burn out.''} (P8)

\subsubsection{Power Sharing Structures} 
While a moderating team members hold the power to execute governing actions (e.g., interaction limits or bans), they also experienced power restrictions. Restrictions typically originate from higher-up governing bodies such as the Technical Steering Committee, but efforts to decentralize and democratize moderation also encourage community members to call review and call out and misjudgments of moderators. 

Technical Steering Committees (TSCs) tend to appear only in larger projects (only 3 of the 10 projects we interviewed formed one -- Apex, Vessel and Grunge), where the sizable number of internal project members calls for \textbf{top-down governance}. In Apex for instance, P13 was blocked from directly removing an internal member because \textit{``once you're a collaborator you can't really be removed''. In order to remove an internal collaborator, \textit{``the Technical Steering Committee needs to vote \dots we [the moderation team] wouldn't [typically] remove a collaborator''} (P13).} 

The TSC shoulders many technical and governing responsibilities, serving as \textit{``the unifying factor''} of the project (P13). But the TSC also exhibits \textit{``a very strong bias towards inaction, by design \dots because \dots making the wrong technical decision is a lot riskier than not making technical decision}.'' 
Finally, the TSC also consists \textit{``a lot of people who are very technical, [so] they don't like dealing with interpersonal issues''} (P13). The combination of limited bandwidth, composition of technical members, and tendency toward inaction means that \textbf{the TSC is slow in approving requests} for actions like removing collaborators. As a result, maintainers of larger projects eventually \textit{``determined that we needed a separate body from the Community Committee and the Technical Steering Committee to handle these [governance actions] because membership on the TSC does not mean you have any idea how to handle a code of conduct report''} (P8), leading to the formation of official moderation teams in project Vessel.

The \textbf{TSC holds powers above the moderation team} (e.g., the ability to remove internal collaborators), and moderators must additionally \textit{``do a weekly report to the TSC about what moderation actions have happened \dots [to adhere to] \dots our governance documentation''} (P8). In addition, moderation teams set up structures to also encourage \textbf{project members to check their judgments} as well, so as to ensure a more democratic distribution of moderating powers:
\begin{quote}
    ``We always invite people to call the other mods to check that we are actually right because we get it wrong. Because otherwise if we wouldn't have rules to follow then it would be, well `this mod didn't like my nose so he banned me' '' (P5)
\end{quote}

\subsubsection{Reporting Mechanisms} \label{reporting}
To support the reporting of misconduct from volunteers, moderators of larger projects set up \textit{``a private moderation repo''} so that \textit{``collaborators ($\sim$500-600ish of them)'}' can \textit{``open issues there to notify the moderation team that `here it's something that \dots you need to look at' ''} (P8). These \textbf{moderation repos for community reporting} works for larger projects because \textit{``for very contentious topics [in] issues and pull requests (which happen occasionally in most projects), someone will notice and surface it even though there's nothing bad yet''}. In addition to providing a centralized place for members to submit reports, this strategy enables moderation team members to \textit{``start subscribing to it and jump really quickly when something happens''}. 


In addition moderation efforts from the community, \textbf{reports to GitHub} constitute another avenue for escalation if moderators don't have the power to edit particular posts or close specific user accounts. For instance, P8 relates how \textit{``There are definitely some blind spots and missing parts \dots certain types of comments you can't edit or delete \dots that's a bit of a problem. We have to contact GitHub if it gets really bad -- but if it gets really bad you just report the user and eventually all their stuff gets deleted because GitHub just deletes the user}". Spammers who occasionally attacked the mid-sized project Hyundai were dealt with in a  similar way: \textit{``spammers that come with things (political) that doesn't have anything to do with Hyundai, they occur twice a year and we have to delete/close issues. They are also reported to GitHub, who may close their accounts''}. 

\begin{table*}[h]
\begin{tabular}{|p{0.13\linewidth} | p{0.54\linewidth} | p{0.25\linewidth} |}
\hline
\textbf{Moderation Strategy} & \textbf{Description} & \textbf{Example Actions} \\ \hline
Punitive & 
Reactive measures taken to eliminate harmful content and prohibit interactions that cause rapid and excessive negative engagements. Used in situations when someone acts in a clearly outlawed manner or activities that cause high levels of community response & Hiding/deleting comments, bans, interaction limits, locking conversations, calling out bad behavior. \\ \hline
Mediations & Diplomatic interventions taken to resolve small-scale misunderstandings and agreements. Used for disagreements between a small number of (usually internal) contributors. & Correcting misunderstandings, forming negotiations. \\ \hline
\multirow{2}{*}{Preventative} & Inhibitory: Precautionary measures used to prevent the development and further escalations of conflicts. Used in situations that maintainers perceive to have the potential to escalate, such as expressions of indirect hostility, inside jokes, belittling comments & Issuing warnings, calling out behavior that are perceived to have potential to escalate. \\ \cline{2-3} 
 & Proactive: Setting up rules and workflows to avoid the repetition of similar mistakes and future user/contributor frustrations. Used after repeated offenses. & Setting up private moderation repos, codes of conduct, linters, templates, topic-specific channels \\ \hline
Reformative & Educational approaches to rehabilitate misbehaviors and set up acceptable standards. Used after unintended neglect of rules or repeated violations by multiple members. & Offering explanations, polite admonishment. \\ \hline
\end{tabular}
\caption{Summary of Moderation Strategies}
    \label{tab:tab2}
\end{table*}
\subsection{Moderation Strategies} \label{2}
In an ideal world, maintainers should not have to monitor and respond to negative interactions. But despite their best intentions, contributors did end up engaging in heated conversations that escalate quickly out of control. When such unexpected situations occurred, maintainers \textbf{reacted} by utilizing a set of existing tools on GitHub to help limit, de-escalate or remove the interaction. But sometimes it takes more in-depth interventions to resolve a conflict, in which case maintainers and moderators performed the role of the conciliator to \textbf{mediate} the dispute. Fortunately, many misbehaviors can be anticipated and prevented once moderators have witnessed and intervened in similar incidents. In such instances, moderators took \textbf{preventative} actions to deter further escalations and avoid future mistakes or \textbf{reformative} strategies so that newcomers to the project can distinguish the acceptable behaviors from the inappropriate. Once established, norms guided contributors toward more productive, healthy and efficient interactions. Table \ref{tab:tab2} shows definitions of the moderation strategies we uncovered, and example actions associated with each strategy. 
\subsubsection{Punitive Strategies} \label{2.1}
Punitive strategies consist of reactive moderating actions such as content removal, bans, locking of conversations, or strict enforcement of codes of conduct guidelines to eliminate harmful content and disruptive behaviors. These were usually taken immediately after severe situations such as unexpected debates and outbursts, so as to limit the impact of inappropriate actions and prevent further escalations of conflict. 

When content removal was sufficient to conclude and archive an exchange, moderators simply hid or deleted comments. P3 of Apex related his latest \textbf{preference for hiding comments} over deletion since GitHub introduced public deletion receipts : \textit{``I don't delete comments anymore because GitHub leaves a record that you've done it \dots because of that it's more effective to hide them all as abusive or off-topic''}.

Unlike deletion (which now leaves a public trail of delete receipts), folding the content via hiding offered transparency, which was found to 1.) improve legitimacy and accountability 2.) increase perceived consistency and 3.) prevent confusion and frustration \cite{Jiang2022-lg}: \textit{``People can still read it (which sucks), but then there's not an illusion of censorship, which is worse than people reading the content, but not as good as the content being erased. ''} (P3). Prior to public deletion receipts, \textit{``if there was a hugely toxic exchange that was irrelevant to the issue, I could sum it up and then delete all the comments and nobody had to see that toxic exchange had happened''} (P3).

Moderators found it was crucial to \textbf{enforce existing rules} to maintain a healthy and supportive environment. In the case of political spam in Hyundai, P6 recounted having to delete and close issues, as well as the report the accounts to GitHub. To clearly outline desired behaviors, it can be helpful to have a \textit{``code of conduct, and being open about enforcing it helps a lot, because people know what they're getting if they go that way…acceptable behavior is pretty much laid out [there]''} (P5). P5 additionally emphasized the importance of invoking existing rules: 

\begin{quote}
``We have the moderation team to enforce this; we want to be constructive at all times. We do not accept people harassing other people or calling them names or generally being negative, it's rather frowned upon. Basically we criticize the code not the person: be constructive, be on the point.'' 
\end{quote}

Moderators also called out clearly toxic behaviors that were not yet explicitly delineated in existing rules. For instance, within the project of popular language Grunge: \textit{``we \textbf{call out bad behavior} when we see it.''} (P5). These concerns were raised directly on GitHub or through other media: P9 and his team members on Vessel \textit{``call out bad behavior by sending screenshots over the team's Slack''} while P13 and his team on Apex post in a moderation repo to encourage accountability \textit{``You open an issue in moderation repo, so they see that you're aware of it \dots and often that's enough to get them to de-escalate and no other people are watching.''}  

\textbf{Reactive approaches require quick responses}, since escalations tend to unravel quickly \textit{``Either we don't notice it, or we say `hey it's banned' or `work it out' \dots But \dots sometimes it's a bad thing to catch it like one day too late, and at that point it's too late.''} (P13). Should disagreements develop into more heated debates, moderators would institute temporary bans: \textit{``There will sometimes be very heated discussions, we may institute a one-, or even in some cases the seven-day ban, so they can cool off and then come back, refreshed, hopefully}.'' (P5).

\subsubsection{Mediations}\label{mediations}
\textit{``In an OS community, the implicit foundation of it is that all contributions are valid, or that everybody has an equal stake in doing something''} (P11). But disagreements occurred when maintainers and contributors have mismatched expectations for the future state of the software \cite{Geiger2021-ay}. During such interpersonal conflicts, it fell to moderators to hear out all perspectives and mediate underlying conflicts to resolve disagreement and limit the development of toxic behaviors. 

Mediation involves \textbf{communicating with multiple parties} involved in a conflict (individually or as a group) to resolve any misunderstandings or negotiate any conflicting objectives when collaborating on a decision in the project. P13 described how one party engaged in a conflict sought out moderators to mediate situations:  \textit{``You find the moderator that is respected by both parties that are involved in the conflict. \dots Then you talk to them, if they're nice they usually agree to facilitate things. You get them to hear both sides, they take it from there.''} And having conducted mediations himself, P13 elaborated on the sequential process of mediations. To start off, the moderator speaks with individuals from both sides of the conflict: \textit{``you just talk to the sides, 
\dots you try to figure out the conflict, you try to get them to see the other person's perspective''}. In some cases someone actually did commit a wrongdoing or misconduct: \textit{``Sometimes there is a clear person who is right in the conflict \dots  usually the other party will either admit that or dig in}.'' But more likely it's just a miscommunication:  \textit{``Often there is not [someone in the wrong], it's just like a misunderstanding, and just getting people to see the misunderstanding and the other person's perspective is usually enough''}.

P13 of project Apex also recounted approaching mediation by \textbf{giving all parties the benefit of doubt}: \textit{``Most of these people are good people and good engineers, and there is very little malice in the project. Just assuming good faith and trying to approach it from a point of like `these are reasonable, decent human beings' is often sufficient, in terms of figuring out the right side.''} Meanwhile, P14 of a smaller project found mediation to be \textbf{a negotiatory task}: \textit{``It's all about negotiation. You talk with engineer A, you tell them what you don't like. You try to talk with engineer B, you try to see if what engineer A is proposing will work with engineer B, and you try to come up with a tradeoff.''}

Some maintainers were happy to act as an intermediary from the beginning. For instance P1 related how \textit{``I would rather be [a] middleman than to call out anyone for toxicity''} while P14 helped his contributors ask for clarifications: \textit{``Sometimes people can come to me and say: `I read this, not sure how to take it, if it's personal or something'. Usually I know all interested parties and I try to ask the reviewer to rephrase the message, to clarify it.''} But founders of more mature projects like JaguarAPI were not as comfortable with mediating - \textit{``I'm trying to mediate, which is strange because that's not something I would normally do. I wouldn't normally engage in an aggressive conversation''} (P4). But due to hypervisibility of the project and obligations to protect community members, P4 wound up learning how to learn anyway: \textit{``I feel like I have to protect the community and the people that are around there, around my family. So I end up having to stop whoever is kind of harassing us''} .

\subsubsection{Preventative Strategies} 
Mediations and punitive strategies describe ways that moderators react to conflicts of different scales. While these techniques can be taught and directly performed by any new moderator, it takes more experience to anticipate and prevent budding or future disputes. Kiesler et al. presented ways to limit the \textbf{impacts} of misbehaviors as well as the \textbf{performance} of bad behaviors in their meta-analysis \cite{noauthor_2012-sr}. Below we categorized these two types of strategies as \textbf{inhibitory and proactive preventions}, where moderators used the former to prevent escalations the latter to proactively set up workflows to prevent frustrations and ensure conformity to standards.

\paragraph{Inhibitory Preventions} \label{inhibitory}
Not all conflicts end up escalating into full-blown arguments between contributors, and most of the time it was up to human moderators to predict the onset of harmful behaviors. 
Inhibitory preventions involve \textbf{warning-based, reproachful techniques} that moderators leverage to target indirectly hostile behaviors (e.g. inappropriate jokes, passive aggressive behaviors), so as \textbf{to limit harm} and avoid further escalations. 
The indirectly hostile behavior in open source projects is analogous to the concept of ``toxicity elicitation'' in online text-based communities \cite{Xia2020-zd}, which are comments or behaviors that elicit high toxic responses, but doesn't necessarily contain toxic language itself. 
The preventative actions targeting these behaviors included monitoring conversations, calling out and correcting misbehaviors, or issuing warnings.

Passive aggressive behaviors were a classic example of indirect hostility that participants brought up, and P5 of Grunge recounts how \textit{``we always stop this. You have to \textbf{nip it in the bud}, because people new to the language come there to ask questions, that's always a delicate situation''}. To reduce the chances of newcomers dropping off, \textit{``we're extra careful there to protect those people from know-it-alls and people who just ooze negativity''} (P5). P11 from Silverback similarly practiced the firm enforcement of rules to prevent escalations: \textit{``you just firmly enforce it, and that itself creates a good culture because you nip these things in the bud. You don't let them escalate out''} (P11). In the absence of existing rules, moderators issued preventative warnings to de-escalate situations. \textit{``Other than bans, it can even just be a proverbial slap on the wrist. We call out bad behavior and if they fix it directly then that's totally okay, we appreciate that not everyone is at their best''} (P5).

Even though these comments were not as outright and blatantly harmful, they did contribute to the normalization of hidden hostility: 
\begin{quote}
    ``The problem isn't the offensive or toxic comments, that's not actually the issue. It's not actually a problem that someone is entitled, in a comment directly. It's the knock on effects of that comment, it's that other people will see that and think it's okay to behave that way, it's that other people will feel more entitled because they've seen entitlement be normalized.'' (P3)
\end{quote}
\paragraph{Proactive Preventions} \label{proactive}
After observing repeated instances of misconduct, moderators \textbf{proactively established rules and workflow standards} to avoid the repetition of similar mistakes, minimize the amount of harm that bad actors can perform, and guide new contributors toward desired standards and practices, which has been found as an issue for newcomers \cite{Geiger2021-ay}. Specific structures include codes of conduct, private moderation repos, formatting linters, templates that help contributors better frame their questions and suggestions, or channels for organizing existing answers. While these structures did not directly take place after an offense, their presence created structures for support and information dissemination, thereby minimizing questions and issues raised from users and contributors.

In the case of P8 from Apex, an entire moderation team was set up in reaction to a conflict: ``The moderation team is set up in reaction to Apex botching \dots [a situation]. It was a public relations fiasco\textit{.''} A team member P3 also described how besides moderation teams, contributors also set up codes of conduct after instances of conflict: ``anyone who has run into the need for moderation or codes of conduct, is going to be very quick to implement it in a community they enter or create''.

To minimize the edits that maintainers need to make to contributors' submissions, \textbf{templates} helped to list out necessary components to include in new pull requests or issues: \textit{``In the repository, in the topmost folder, there is a Contributing document that says: your pull request should have this title, commit messages should be in this format''} (P3) or assist users in drafting issues \textit{``I added a load of information to the template and a lot of requisites to ask people to build a very simple example of what is it that you want.''} (P4).

\subsubsection{Reformative Strategies}\label{reformative}
Not all acts of misconduct are borne from malicious intentions \cite{Jiang2022-lg}. Sometimes when moderators observed repeated instances of misbehaviors, they employed a more nurturing and reformative approach that doesn't castigate contributors for unintentional offenses.
Unlike the punitive or preventative strategies that remove a member's content or right to interactions, reformative techniques are more educational and gentle, consisting of actions like polite admonishments or providing explanations. 
Over the long term, artifacts from reformative approaches (e.g. explanations) benefit the community by establishing acceptable behavioral standards, even if they take some time for communities to adopt.
By offering benefits such as transparency and a way to establish new norms for subsequent community members, reformative approaches garnered increased advocacy from researchers in recent years \cite{Myers_West2018-hu,Jhaver2019-zo}.

Similarly, \textbf{reformative strategies were well received} among open source practitioners as well. P3 of Apex related positive feedback from his community:
\textit{``The polite admonishment (when I word it eloquently enough) tends to gather lots of heart and thumb-up emoji reactions, and the person will either apologize or just dip out and be quiet. So it's the most effective form of response.''} In a newer project, P11 redirected a raised issue to demonstrate a more efficient response for typos: \textit{``Thanks for point this out. However, instead of raising this as an issue, if you ever see small typos, please feel free to just put in a pull request to fix them}". However, \textbf{providing explanations is a nontrivial amount of work}, so sometimes maintainers fall back on preventative strategies: \textit{``I don't always have the energy for that so sometimes I'm hostile back \dots sometimes biting comments in response are effective, at the cost of other people seeing me as a jerk, but it still establishes that behavior is not acceptable''} (P3).

One side effect of politely admonishing community members is the \textbf{potential loss of a contributor}, but often that risk is outweighed by the knock-on effects of unaddressed misbehaviors: 
\begin{quote}
Establishing what behavior is acceptable and what is not \dots [it] is performative - it's showing everyone else in the arena that that behavior is not okay, even if that means that person is not going to improve. And while it's always preferred to rehabilitate someone, or convince them to re-evaluate\dots I'd rather lose a person forever from the community than have the rest of the community see toxic behavior go unchallenged.
\end{quote}

In addition to polite admonishment, P2 of a new differential privacy library showed how reformative actions also offer \textbf{explanations of newly established norms and practices}: \textit{``New pipelines are introduced through meetings, and introducers explain why they're better, they are then more accepted by contributors after explanation''}.
Some of these standards took a period of transition for some communities to adopt, demonstrating a case of normative conflict identified from \cite{Filippova2015-yi}: 
\begin{quote}
    "When a community starts moderating it's overwhelming for a while\dots The goal is to get everyone to be as open and tolerant and respectful as possible and that goal is not \dots most efficiently achieved by immediately jumping to a list of all the things that are potentially a problem \dots each medium has to get there on its own time, in its own way so the norms can be established and everyone can accept them." (P3)
\end{quote}
However, once communities adopted a good practice, they grew to appreciate it over the long run: \textit{``And then norms are established; they know that it's safe to admonish newcomers to behave that way. And the incidence of reports just plummets. People just don't screw up when they know what the norms are supposed to be. We'll get through the transition period and the new norms will just be the way it is, and everyone will be horrified that it used to be worse.''} (P3). Among collaborators, such \textbf{adoption frictions} were usually mitigated by group meetings and discussion: \textit{``New pipelines are introduced through meetings, and introducers explain why they're better, they are then more accepted by contributors after explanation.''} (P2).

\subsection{Automation of Moderation}
Most of the interactions on social coding platforms are text-based, making them well-suited for automation when compared to their social media counterparts \cite{Jiang2019-tr}. As a result, bots and GitHub Actions easily leverage the available repo artifacts to facilitate various workflows and protection mechanisms for their projects \cite{Wessel2021-pi}. \del{7 of our 14}\add{Half of our} participants mentioned using or considering automated tools to facilitate community moderation. While most of them did not currently have moderation tools set up on their repos, Hyundai had a positive experience using the Sentiment Bot, and Silverback had installed the alex bot, which detects instances of ``gender favoring, polarizing, race related, or other unequal phrasing in text'' \cite{Wormer2015-ff}. 

However, our interviewees perceived current bots to be inadequate for conducting moderation beyond simple reactive warnings. Moderators reported that community members can view automated moderation tools as over-censoring and policing forces that threaten their freedom of speech, especially due to their tendencies for false triggers. Furthermore, the more subtle forms of misbehaviors found in the professional space of software development (such as those covered in Section \ref{subtle} \nameref{subtle}) are difficult to anticipate by the language models' underlying moderation tools. Meanwhile, the tools used for moderation are seldom adapted to the development context and lack access to cross-platform information, increasing the chance for false alarms. Finally, the absence of customization options for privacy and notifications breached social boundaries between users, contributors and maintainers by exposing deletions and callouts to the public and unnecessarily demanding maintainers' attention with excessive and overly public notifications. So despite the potential for bots to perform automated moderation on the behalf of maintainers, many of our participants expressed concerns and adoption frictions. Below we highlight some of these existing tensions as well as maintainers' stance on the utility and impact of moderation bots. 

\subsubsection{Automated Moderation Breeds Over-Censorship} \label{3.3}
In public online spaces, the right to \textbf{free individual expression inevitably trades off with concerns of wellbeing and public safety}  \cite{Grover2019-ep}, and our interview participants perceived the potential for automation to lead to over-censorship. Free speech has long-standing associations with source code, and the metaphor was leveraged by early supporters of F/OSS to protect the right to use, modify and distribute software \cite{Coleman2009-ru}. As a result, open source communities have strongly embraced and valued the right to free speech. However, in volunteer-based development contexts, it is also important to foster a safe space that welcomes contributions from everyone, especially given the limited diversity and inclusion in modern open source communities \cite{Li2021-rx}; P8 of Apex points out the shortcoming \textit{``As with a lot of organizations \dots we struggle with representation.''} But as Gibson has found, moderators use punitive strategies more within safe spaces \cite{Gibson2019-zn}, which leads to a sense of over-censorship. Teammate P3 described the two opposing value systems: 
\begin{quote}
    ``There's a spectrum of how much we want to modify our language to avoid offending people. And the folks who generally resist political correctness are the ones who are on the side of saying, `\textit{I'm going to say whatever I want, if you're offended that's your problem}' and I think that sucks and I don't want to be there. But the other extreme is problematic too because it damages the message by turning into policing.''
\end{quote}

\textbf{Maintainers delicately balanced this friction} between individual contributors' desire for free speech and the community's need to create a welcoming space for female, LGBTQ, and other underrepresented groups \cite{Li2021-rx,Vasilescu2015-tq}. Depending on contextual needs, maintainers balanced the freedom of expression for contributors with  broader project goals of promoting an respectful and inclusive community, or as P14 eloquently expressed: \textit{``the tradeoff between this [moderation] and freedom. You have to balance, don't want to restrict people, but you also want everybody to play nice.''}

While some maintainers struggled with the tradeoff between free expression and enforced civility, P10  (founder of Silverback) expressed \textbf{active resistance against demands for free speech}: \textit{``This is a slippery slope, [as if] we're going to `lose all of our words' and `it's going to be 1984' and `we can't express ourselves'.''} To assert that automations are not powerful enough to suppress our creativity and right to speech, P10 challenged users/contributors who complained to test out AlexJS, telling them that: \textit{``if you lose more than like 15 words in a year red we can rediscuss this. But \dots you're smart enough and creative enough, and the language is large enough, that you're going to be just fine''}.

Most of the automation assistance our participants considered focused on language use, and less so on other forms of misbehavior. However, excessive attention to and moderation of language misuse also \textbf{derailed conversations topics away from the development of software} - \textit{``even the people who were philosophically aligned with the idea of avoiding gender words were still irritated that the normal topic of the channel was distracted or disrupted by that conversation so frequently''} (P3). Such concerns are another set of nuances that bots would have a hard time taking into account: \textit{``if you get a tool like that -- it can legitimately be seen as being too pedantic or too tightly wound about certain words. And [if] the culture of that community isn't ready for that yet, then that's worse than not saying anything''} (P3).

\subsubsection{Perceived Technical Limitations of Automating Moderation}
Participants perceived that moderation bots deployed on GitHub doubly suffered from context specificity due to 1.) situation-specific nuances that are difficult for current tools to pick up on and 2.) technical terms used in software development environments.
On the one hand, a lack of nuanced understanding of situational contexts made it difficult for models to detect new and more subtle variations of misbehaviors. 
On the other hand, the underlying language models lacked contextual sensitivity to technical terms, triggering false positives that require additional human labor to review. 
The combination of such shortcomings caused hesitation in delegating moderation responsibilities to automation tools.

\paragraph{Inability to Anticipate False Negatives} Human collaborators can easily retrieve information distributed across platforms by toggling between them, but most automation tools can only access a single deployment context or platform: ``a lot of the mediums [where] they have discussions are not in public -- the bot wouldn't have access'' (P13). Without multiplatform contextual clues, bots failed to pick up on interpersonal relationships or intended meanings from a working environment, and \textit{``even inside the discussion there is a lot of background [and] the bot would have a very hard time to figure it out''} (P13). For instance, P6 of Hyundai pointed out that \textit{``when people are passive aggressive, the bot cannot understand that, and it's better to interact with a human''}. Hence, much like the moderators of social media contexts, \textbf{open source moderators prefer human-reviewed} decisions \textit{``for the interpersonal conflict inside of projects [since] it would be impossible''} for automated assistance of moderation to work \cite{Seering_Wang_Yoon_Kaufman_2019}. P3 also shares the stance that human should be involved in moderating decisions: \textit{``If it was just maintainers that see it: fine, I can make a judgment call''.} 

\paragraph{Context Specificity Raises False Positives } General-purpose sentiment analysis models have struggle to pick up on connotations of context-dependent terms, causing learning-based models to \textbf{falsely trigger} \textbf{on common software engineering terms} that carry negative denotations when used in everyday contexts. Consequently, maintainers had to manually review instances of such false positives triggers to ensure accuracy, exacerbating their already limited bandwidth. 
For instance, P3 introduced how AlexJS could offend individuals by aggressively flagging words with slightly negative connotations: \textit{``AlexJS \dots ended up \dots [being similar to] the archeology conference where they couldn't say the word bone, because some software I flagged it as offensive}.'' In project Silerback, P10 also observed how \textit{``Alex triggers on `master' because of [the upstream dependency] Vessel''}  because project Vessel had not yet renamed their `\textit{master}' branch as `\textit{main}'.

One particular example of contextual information missing from detection models was a \textbf{contributor's} \textbf{primary language}. When someone's native tongue is not English, their comments can accidentally trigger bot reactions by unintentionally using phrases that carry negative connotations and innuendos: \textit{``English not being a first language may affect them.''} (P8). Li et al. has highlighted how moderators must use intuition (i.e. guessing) to discern between behaviors needing intervention from unintentional offenses caused by language differences \cite{Li2021-rx}. 

Participants also worried about how bots will treat \textbf{self-directed anger}. In one instance, the founder of project Hyundai \textit{``was answering a question and said `Oh yes, don't worry about that feature, it was a rubbish feature and we already fixed it' and the bot was triggered.''} (P6). Likewise, P10 \textit{``worries about the use of negative language in code due to personal experience writing code (mostly self-directed).''} While P6 was able to find humor in the situation, \textit{``Sometimes the Sentiment Bot flags not so aggressive phrases and it's a funny occurrence''} other instances may be more frustrating, especially if the triggering words are frequently used.

Most maintainers hold the opinion that false positives cause harm, especially if they add noisy information: \textit{``false positives become a big problem, especially because they're a distraction''} (P5). But the potential for false negatives to cause disruptions depend on context: \textit{``Sometimes false positives are acceptable, better than missing something \dots but there's also sometimes where false anything is not acceptable and it's better to say nothing, than to have a false result.''} (P3). Too many \textbf{false triggers can numb maintainers' responses to warnings} (a behavior consistent with the findings of Wessel et al. \cite{Wessel2021-pi}), perceiving them as noise instead and thereafter ignoring them altogether: \textit{``We have something called a stale bot \dots it periodically will just put comments on tickets and send emails, which is not bad. But for whatever reason we've learned to ignore it sometimes.''} (P10).

\subsubsection{Customizations and Boundaries}
Participants reported strong needs to tweak and customize the tooling based on specific project needs. In existing automation tools, the \textbf{lack of personalization options} harmed adoption rates. For instance, the absence of options for notification settings caused information overload and fatigue for maintainers, especially given the possiblity of the abovementioned false positive triggers .

In Sections \ref{self-moderation} and \ref{reporting} we discussed the distribution of moderation work to volunteer contributors. P3 brought up an in-house GitHub feature that supports this volunteer reporting framework -- \textit{``GitHub has this reporting facility that I don't think too many people know to turn on. It allows arbitrary users to say this is a problem, pay attention''}. Unfortunately the feature lacks \textbf{notification settings}: \textit{``In [Apex] we actually turned that off (it was on for a while) because you can only turn it on for the whole org or none of it}.'' 

\textbf{Privacy} is another setting that maintainers wanted to configure. While transparency has been found to improve collaboration in open source \cite{Dabbish2012-yi}, not all maintainers are ready \textit{``to be that transparent, and that direct''}, and P3 is of the mind set that completely transparent configurations are \textit{``going to have consequences that those folks didn't anticipate and that a private system allows for more bias''}. However, they also conceded later that privacy ``also allows \dots for a more refined response, hence it is of importance to have the agency to configure for private notifications: \textit{``Having it surface just so I notice it \dots [it should also be that] I can also tweak how sensitive it is, instead of having a default setting''}.

There exist yet other bots that are intended to lessen the burden of maintainers, but have instead crossed a social boundary that maintainers were not entirely comfortable with. For example, P3 worried about how the automatic closing of issues deprioritized the time of community contributors: \textit{``I don't use Probot at all primarily [because] most of the usage of it I've seen has been programmatically closing issues (like stale issues) and I think that's insanely user hostile \dots prioritizing maintainer time over the feelings of users and I think that's not a good trade off.''} In a similar vein, P1 also claimed he \textit{``would not consider anything that'd directly communicate with the contributor, because we value every single one of them''} and allowing direct communication between automated moderation tools and contributors could risk offending and losing valuable community members. 


\subsubsection{Anticipated Role of Bots in Moderation}
Perhaps due to the shortcomings outlined above, maintainers for repos of various sizes indicated that \textbf{projects should not (solely) depend on automation for moderation}. As a moderator of the popular project Grunge, P5 thought its presence would be extraneous - \textit{``the best thing it could do was alert as to situations that may arise. But then again, people already do that''}. As the creator of the more nascent Silverback project, P11 also thought that his community - \textit{``should never need it, other than catching slip-ups''} because \textit{``if we're relying on the bot to solve moderation problems, we've gone so far off course''}.

Fortunately, the future for bot adoption is not entirely dismal. P8 expressed appreciation for the \textbf{depersonalized nature of automated interventions}, suggesting that it may have a place in initiating interpersonal interactions such as mediations \textit{``it's nice that the tool depersonalizes the intervention''}. But like bots on any platform, user abuse is a possibility: \textit{``as soon as a repo has that system on it \dots a bunch of people are going to go brigade it and just drop every offensive word they can think \dots just to see how much they can respond'' }(P3).

Additionally, our participants considered scenarios where moderation bots can be leveraged to execute some of the \nameref{2} -- Table \ref{tab:tab3} overviews some ways that bots can support moderation in the future. For situations needing immediate response such as warnings that are administered through \nameref{2.1}, P12 recalled an instance of a demanding user where a bot could have intervened. The user had commented: `\textit{``THIS BUG HAS BEEN OPEN FOR A YEAR, WHERE IS THE FIX AT' in all caps -- I can definitely see using it [a moderation bot] for something like that.''} P4 also contemplated a situation where the Sentiment Bot could have taken the frontline, reactive work of moderation: \textit{``if it was able to take a lot of those first conversations I think that will be very useful''}. P8 imagined that such a tool could help self-moderation by alerting well-intentioned commenters when they accidentally make a mistake: \textit{``This is gonna be good for people who are good faith commenters, it's not gonna be effective for the trolls''}. In communities where malpractices were pervasive, P14 imagined that moderation bots could help facilitate reform: \textit{``If I see this type of behavior becoming a bad practice in the team/overall community, I would definitely consider doing [adopting] something like that''}. 

\begin{table*}
    \centering
    \begin{tabular}{|p{0.15\linewidth} | p{0.25\linewidth} | p{0.25\linewidth} | p{0.25\linewidth}|}
    \hline
      & \multicolumn{1}{l|}{\textbf{Self moderation}} & \multicolumn{1}{l|}{\textbf{Volunteer Moderators}} & \textbf{Moderation Teams} \\ \hline
      \textbf{Punitive} & \multicolumn{1}{p{0.25\linewidth}|}{Current bot interventions are suitable for reactive self-moderation, but customizations, contextual sensitivity and higher accuracy can increase usage.} & \multicolumn{1}{p{0.25\linewidth}|}{Current volunteer moderators often experience false positives triggers from moderation bots, customized notifications, increased accuracy and contextual sensitivity can encourage adoption.} & Bot interventions can help improve the efficiency of content moderation in large projects with moderation teams. There are opportunities for bots to help team members to make collaborative decisions or onboard new members. \\ \hline
      \textbf{Mediations} & \multicolumn{1}{p{0.25\linewidth}|}{N/A (Conflicts involving mediation have usually escalated beyond self-moderation.)} & \multicolumn{1}{p{0.25\linewidth}|}{Bots can ask for clarifications on behalf of a contributor (acting as a mediator) in place of moderators.} & Bots can help depersonalize mediations but there exists room for improvement in detecting situations that are in need of mediations in large projects. \\ \hline
      \textbf{Preventative} & \multicolumn{3}{p{0.75\linewidth}|}{Inhibitory: There are opportunities for detecting instances of potential toxicity such as indirect hostility that could develop into more serious conflicts (e.g. passive aggressiveness, inside jokes, minor transgressions). Proactive: Bots can provide suggestions of improved workflows after observing repeated mistakes and unconformity to existing standards.} \\ \hline
      \textbf{Reformative} & \multicolumn{3}{p{0.75\linewidth}|}{Bots can help enforce template use and surface rules, community guidelines and codes of conduct to writers when they are composing a potentially harmful comment.} \\ \hline
    \end{tabular}
    \caption{Design Recommendations: how automation may support moderatiom structures and strategies}
    \label{tab:tab3}
\end{table*}

\section{Discussion}
Through our examination of moderation norms and practices among communities of various sizes, we found a diverse set of structures and practices that maintainers leverage to manage and prevent conflicts. While self-moderation and volunteer-based moderation have pervaded and been well-studied in neighboring communities such as Wikipedia \cite{Forte2009-ze} and Stack Overflow \cite{Cheriyan2020-qp}, we found that moderation required a different set of strategies and in the case of larger projects, more formal structures such as moderation teams. We also discovered that there are still many gaps in the forms of moderation assistance that bots can offer, both in terms of whom they serve and the type of moderation strategies they automate. Inspired by some speculations of our participants, we present below a comparison between the moderation structures, strategies and opportunities for automation in open source versus other platforms, as well as some design recommendations to help guide the future of automation tools for moderation.

\subsection{Moderation in open source versus other platforms}

\subsubsection{Moderated Content}
In terms of content, prior works on content moderation in social media largely documented the presence of more explicit forms of misbehaviors such as the infamous triad of \textit{``flaming, spamming and virtual rape,''} among other forms of inappropriate content such as hate speech, insults or harassment \cite{Seering_Wang_Yoon_Kaufman_2019, Jiang2019-tr, Gilbert_2017, Jhaver2018-ln}. In our discussions with practitioners on GitHub, we gathered that moderators also watched out for more borderline actions such as technical disagreements or resistance to new norms, which may not be as immediately apparent. The evidence of such subtle forms of disputes mean that moderators are more likely to leverage \nameref{mediations} as an approach to conflicts between contributors. Another implication is that automated tools powered by language models are unlikely to detect these less obvious misbehaviors, because not only are they subtle, but they also tend to be situational and technical -- and therefore highly context-dependent.

\subsubsection{Structures and Roles}
While prior works discuss the potentials of self-moderation in community-based platforms such as Facebook groups, Wikipedia and Reddit \cite{Seering_2020, bozarth2023wisdom}, most considered moderation to be a community-level effort where platform peers helped one another moderate, similar to the volunteer moderation that we attribute to community members in this study. 
Among our participants, self-moderation was considered an individually-initiated action where contributors self-monitor and edit their own content. Such behaviors are likely to benefit from automated assistance, as presented in Table \ref{tab:tab3}. Our terminology was consistent with one other study on YouTube \cite{ma2021advertiser}, while another investigation of subreddits called the phenomenon self-censorship \cite{Gibson2019-zn}. 

Many of the communities that practice the community-level self-moderation rely on volunteers to conduct moderation, as opposed to more centralized models of corporate moderation \cite{Seering_Wang_Yoon_Kaufman_2019, Jiang2019-tr}. However, past work suggest that the reliance online volunteers to conduct moderation labor may be exploitative, meriting re-examinations from an ethical perspective \cite{li2022ethical}.
Our results revealed that moderators in OSS shared governing powers with higher-up authorities such as the TSC, as well as with the community members more broadly. Prior work suggest that these mechanisms for distributing power across multiple hierarchical levels is beneficial and expected for larger projects,
by arguing that 1.) power limitations on moderators can increase the perceived legitimacy of their decisions \cite{noauthor_2012-sr}, and 2.) the growth of communities increase the decentralization of moderation on platforms such as Wikipedia \cite{Forte2009-ze}. The establishment of formal structures (such as the moderation teams we introduced in section \ref{formal}) have been found to improve the communication of norms to newcomers \cite{Forte2009-ze}, perhaps by increasing the usage of actions such as \nameref{reformative}.

\subsubsection{Moderation Strategies}
Punitive actions such as hiding or deletion of content as well as the banning and calling out of rule-breaking behaviors resemble much of the organizing actions found on Reddit, Discord and Twitter \cite{Gilbert_2017, Jiang2019-tr, Jhaver2018-ln}, and we found evidence that such strategies are transferable to the OSS context. Similarly, inhibitory warnings used for preventing conflict resemble norm-setting practices adopted by moderators in Wikipedia, Facebook, Twitch, as well as Reddit \cite{forte2008scaling, Seering_Wang_Yoon_Kaufman_2019}. The transition from inhibitory warnings to punitive actions reflects Ostrom's fifth design principle of graduated sanctions \cite{ostrom2000collective}; and though our participants did not explicitly discuss such escalations, we encourage future work to more closely examine its prevalence in OSS contexts. Ferreira et. al. \cite{Ferreira2021-id} advocated for both proactive and reactive (or punitive) approaches for addressing known issues and conducting damage control, and our findings provide evidence of moderators employing such strategies in practice. Finally, mediation was a strategy that was almost never observed among extant literature, except in Wikipedia, perhaps of its similarities with open source as a collaborative peer-production platform \cite{Billings2010-cy}. 

\subsubsection{Usage and Perception of Automation}
Prior works on Wikipedia have found semi- and full-automated tools valuable in providing moderators with an information infrastructure that connected editors from a decentralized network that facilitated valuation, negotiation, and administration, thereby enabling new moderating actions independent of existing norms \cite{Geiger2010-zh}. For open source, Ferreira et. al. anticipated the deployment of similarly automated assistance for moderation \cite{Ferreira2021-id}, yet many others critiqued that existing toxicity detectors are not yet tailored enough for the software engineering context \cite{Ferreira2021-id, Ahmed2017-wt, Jongeling_Datta_Serebrenik_2015}, which our findings corroborated. Beyond the challenges induced by limited domain adaptation, we additionally uncovered the presence of subtle misbehaviors that may contribute to the inability of such models to anticipate more nuanced situations. Lastly, we also highlighted how the absence of customization options caused maintainers to resist adoption, the implementation of which is made difficult by the lack of transparency in the underlying black box models \cite{Juneja2020-kd, geiger2016open}.

\subsection{Design Implications for Automating Moderation}
\paragraph{Punitive strategy} The first set of strategies we uncovered were employed at the early stages of conflict, these included punitive measures that halted the escalation and removed toxic content. Presently, we found that moderation bots are of most assistance to human contributors in this reactive capacity, i.e., by pointing out cases of rule violations and harmful content so that authors can become aware when they unintentionally compose inappropriate content. However, when bots were tasked with calling out bad behaviors, our participants observed that they were prone to hypersensitivity, causing false positive triggers. Such false alarms do not scale well and negatively impact moderators by contributing to their already overloaded maintenance burdens \cite{Wessel2021-pi,Geiger2021-ay}, making existing tools helpful only for cases of self-moderation. To extend the scope of reactive support toward volunteer moderators and formal moderation teams, further improvements of contextual sensitivity in the underlying sentiment models supporting current moderation bots are needed to enhance understanding of nuances of language used in software engineering to increase accuracy. More customization options can also be incorporated into these tools to increase transparency, explainability and trust among the community.

\paragraph{Mediation Strategy} Once conflicts developed, moderators engaged in different approaches to mitigate and resolve issues, depending on specifics of the situation. When encountering disagreements among small parties of contributors, moderators took mediating actions to reconcile differences. During mediations, moderation bots help facilitate depersonalized interventions between contributors, but further advancements can help moderators detect any disputes that require mediation and ask clarifications from a fellow contributor when one side is uncertain about the presence of conflict or the potentially negative connotations of a comment. 

\paragraph{Preventative Strategy} When contributors engage in more indirect forms of toxicity, such as passive aggressiveness or inappropriate jokes, maintainers leveraged inhibitory preventions to limit the extent of bad behaviors. Bots can support moderators by expanding their detection scope to include such forms of indirect hostility. 
After repeated instances of behavioral mistakes occur among different contributors, moderators proactively established new rules and standards to prevent future violations. To contribute toward proactive preventions, bots can help monitor and detect repeated offenses, identify associated workflows that cause such unconformity, and suggest improvements based on practices observed among other communities.

\paragraph{Reformative Strategy} For mistakes repeated by multiple contributors, moderators took reformative approaches to set up standards and proactively prevented future cases of similar violations by introducing new rules and workflows. To help moderators initiate reformation among the community, automation can be utilized to surface existing guidelines in real time when authors are writing content so as to prevent the public posting of potentially harmful content.

\subsection{Relation between Workflow Automations and User Frustration}
While our study did not set out to find out the types of technical and interpersonal conflicts that lead to toxic and uncivil behaviors in open source, there was an emergent theme that pointed to entitlement and user frustrations, especially among more prominent projects with larger user bases, highlighting the shortage of technical support for users and contributors. These problems usually surfaced when participants discussed the types of strategies or workflows they used or set up (mostly described in Sections \ref{proactive} and \ref{3.3}), many of which were established to reduce the masses of questions or requests. While prior work has touched upon how the time-sensitive and never-ending chore of user support is one of the most emotionally draining tasks for maintainers \cite{Geiger2021-ay,Swarts2019-fn}, little is known about the types of technical complexities and misunderstandings that cause such extensive amounts of frustration. Future work can seek to address this missing link between specific forms of technical (or interpersonal) issues that cause these emotionally-charged conflicts as well as some of the mitigation strategies that maintainers mentioned to us above.
\subsection{Limitations}
Our results indicated three main themes around moderation and the potentials of automation in open source, which we presented in the previous sections. However, despite our efforts to recruit from a diverse group of participants and projects for our interviews (with a particular focus on a variety of project sizes), we do not claim that it is representative of all open source developers. The number of interviews we conducted and the snowball sampling technique both limit the representativeness of our sample. We also focused solely on projects hosted on GitHub, which means that the scope of our theory and results may not generalize to other social coding platforms, such as Bitbucket or GitLab. Furthermore, while it would have been ideal to highlight the experiences and perspectives of more marginalized and underrepresented groups in open source, the scarce availability of such participants did not present us the opportunity -- we encourage future work to explore this gap in our understanding of moderators in OSS.

\section{Conclusion}
In this paper, we examined moderation practices in open source communities by conducting 14 semi-structured interviews with moderators and maintainers. Specifically, we characterized the norms, roles and practices of who performs moderation and how different techniques are employed for various contexts (RQ 1). We further investigated automation tools for moderation and identified concerns against adoption, as well as potential ways that future bots can support different groups of moderators in various capacities. Based on the implications of these results, we presented a set of design recommendations for practitioners and researchers, which can guide the future development of automation tools for moderation.

\begin{acks}
This work was supported by the National Science Foundation (NSF) under Award No. 1939606, 2001851, 2000782, 1952085 and 1952085. 
We are grateful to Allen Yao, Pranav Khadpe, Jim Herbsleb, Christian Kastner, David Widder as well as anonymous reviewers for their crucial input and feedback towards the initial and subsequent drafts of this work. Finally, we would like to thank our participants for offering us their time to share their expertise and insights.
\end{acks}

\bibliographystyle{ACM-Reference-Format}
\bibliography{References}

\end{document}